\documentclass[aip,reprint,superscriptaddress,apl]{revtex4-1}

\usepackage{graphicx}
\usepackage{textcomp}
\usepackage{amsmath}
\usepackage{natbib}
\usepackage{dcolumn}
\usepackage{bm}

\begin{document}
\title{Reaching 10~ms single photon lifetimes for superconducting aluminum cavities}
\author{Matthew Reagor}
\affiliation{Department of Physics and Applied Physics, Yale University, New Haven, Connecticut 06520, USA}
\author{Hanhee Paik}
\affiliation{Department of Physics and Applied Physics, Yale University, New Haven, Connecticut 06520, USA}
\affiliation{Current Address: Raytheon BBN Technologies, Cambridge, MA 02138, USA}
\author{Gianluigi Catelani}
\affiliation{Department of Physics and Applied Physics, Yale University, New Haven, Connecticut 06520, USA}
\affiliation{Current Address: Forschungszentrum J\"ulich, Peter Gr\"unberg Institut (PGI-2), 52425 J\"ulich, Germany}
\author{Luyan Sun}
\affiliation{Department of Physics and Applied Physics, Yale University, New Haven, Connecticut 06520, USA}
\author{Christopher Axline}
\affiliation{Department of Physics and Applied Physics, Yale University, New Haven, Connecticut 06520, USA}
\author{Eric Holland}
\affiliation{Department of Physics and Applied Physics, Yale University, New Haven, Connecticut 06520, USA}
\author{Ioan M. Pop}
\affiliation{Department of Physics and Applied Physics, Yale University, New Haven, Connecticut 06520, USA}
\author{Nicholas A. Masluk}
\affiliation{Department of Physics and Applied Physics, Yale University, New Haven, Connecticut 06520, USA}
\affiliation{Current Address: IBM T.J. Watson Research Center, Yorktown Heights, NY 10598, USA}
\author{Teresa Brecht}
\affiliation{Department of Physics and Applied Physics, Yale University, New Haven, Connecticut 06520, USA}
\author{Luigi Frunzio}
\affiliation{Department of Physics and Applied Physics, Yale University, New Haven, Connecticut 06520, USA}
\author{Michel H. Devoret}
\affiliation{Department of Physics and Applied Physics, Yale University, New Haven, Connecticut 06520, USA}
\author{Leonid Glazman}
\affiliation{Department of Physics and Applied Physics, Yale University, New Haven, Connecticut 06520, USA}
\author{Robert J. Schoelkopf}
\email{robert.schoelkopf@yale.edu}
\affiliation{Department of Physics and Applied Physics, Yale University, New Haven, Connecticut 06520, USA}

\date{\today}
\begin{abstract}
Three-dimensional microwave cavities have recently been combined with superconducting qubits in the circuit quantum electrodynamics (cQED) architecture. These cavities should have less sensitivity to dielectric and conductor losses at surfaces and interfaces, which currently limit the performance of planar resonators. We expect that significantly ($>$10$^{3}$) higher quality factors and longer lifetimes should be achievable for 3D structures. Motivated by this principle, we have reached internal quality factors greater than 0.5$\times$10$^{9}$ and intrinsic lifetimes of 0.01 seconds for multiple aluminum superconducting cavity resonators at single photon energies and millikelvin temperatures. These improvements could enable long lived quantum memories with submicrosecond access times when strongly coupled to superconducting qubits.   
\end{abstract}
\maketitle
In circuit quantum electrodynamics (cQED), microwave resonators protect superconducting qubits from decoherence, suppress spontaneous emission\cite{Houck}, allow for quantum non-demolition measurements\cite{Johnson,Hatridge}, and serve as quantum memories\cite{Mariantoni}. Single photon lifetimes between 10-50~$\mu$s ($Q$$\approx$10$^{6}$) have been achieved in thin film resonators with careful surface preparation and geometrical optimization\cite{Megrant,Barends,Geerlings}. The route toward an optimal geometry also sheds light on the physical mechanisms responsible for damping. Planar resonators with larger features are generally found to be higher quality, which is interpreted as loss dominated by surface elements\cite{Megrant,Barends,Geerlings,Gao,Jonas}, as the relative energy stored in surface defects is inversely proportional to the size of the resonator.

Three dimensional, macroscopic cavity resonators are at the extreme limit of this trend and historically exhibit remarkable lifetimes\cite{Turneaure}.  Progress with superconducting niobium cavities for particle acceleration has led to dwell times of seconds for RF field strengths of 10~MeV/m at 2~K bath temperatures\cite{Padamsee}. At the much lower drive powers corresponding to single-photon excitations, or fields of $\sim$1~$\mu$V/m, storage time in excess of 100~ms has been achieved in three dimensional, niobium Fabry Perot resonators at 51~GHz and 0.8~K\cite{Haroche}, and also in 3D, niobium micromaser cavities at 22~GHz and 0.15~K\cite{Brattke}.  The coupling of superconducting qubits to 3D microwave cavities\cite{Paik} could lead to cQED-type experiments with coherence on these timescales. 

We have set out to construct very high quality microwave cavities ($Q$$\gg$10$^{6}$) in superconducting aluminum while focusing on geometries that may be compatible with single-photon cQED experiments at $\sim$10~GHz and 20~mK. We study two types of waveguide cavities (rectangular and cylindrical) that support a diversity of modes to test the effects of material purity and surface treatment on cavity lifetimes in the quantum regime. We find that pure, chemically etched aluminum produces the best results, with rectangular resonators reaching lifetimes, $\tau_{\mathrm{int}}$=$Q_{\mathrm{int}}$/$\omega$ of 1.2~ms ($Q_{\mathrm{int}}$=6.9$\times$10$^{7}$) and cylindrical resonators as long as 10.4~ms ($Q_{\mathrm{int}}$=7.4$\times$10$^{8}$). Realizing these timescales in cQED experiments is a long-standing goal of the field.

In a 3D cavity without bulk dielectric, there are still two types of loss associated with surface imperfections. First, the metal walls could have an oxide layer with a finite dielectric loss tangent. Second, there can be conductive losses due to a finite real part of the superconductor's RF surface impedance. If we first assume the cavity is solely damped by a surface dielectric layer of thickness $t$ and $Q_{\mathrm{diel}}$=1/$\tan\delta$, it will be limited to an internal quality factor of\cite{Pozar} 
\begin{equation}
	Q_{\mathrm{int,E}} = \frac{Q_{\mathrm{diel}} \int_{V}|E|^{2}dV}{\epsilon_{\mathrm{r}} \int_{S}|E|^{2}dA\times t} = \frac{Q_{\mathrm{diel}} V^{\mathrm{E}}_{\mathrm{eff}}}{tS^{\mathrm{E}}_{\mathrm{eff}}} = \frac{Q_{\mathrm{diel}}}{p_{\mathrm{diel}}}~,
\end{equation}
\begin{figure*}[htp]
\centering
\includegraphics[trim=0cm 1cm 0cm 1cm, clip=true, totalheight=0.20	\textheight]{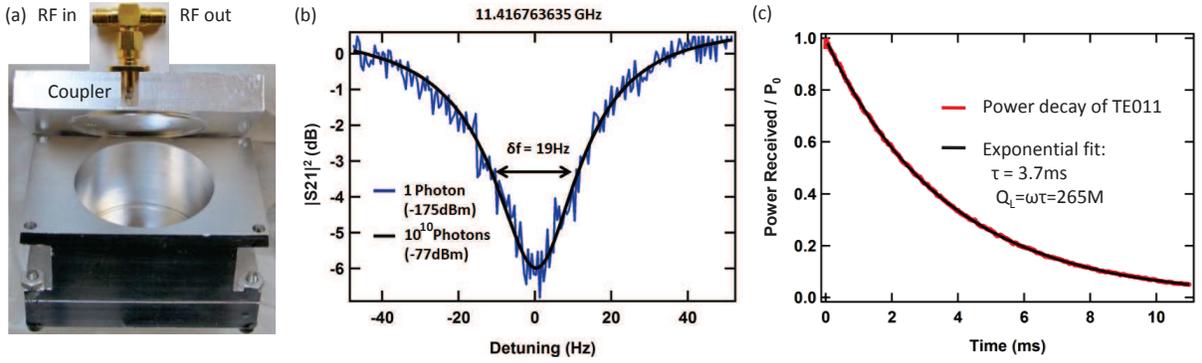}
\caption{(a): Schematic of a cylindrical resonator measured in shunt configuration. (b): Single photon (blue) and high power (black) transmission data of the configuration shown left. The best fit at single photon power is indistinguishable from the high power trace, showing the remarkable power-independence of the TE011 mode, with an internal quality factor of 6$\times$10$^{8}$. (c): Power decay measurements yield a phase-insensitive loaded quality factor $Q_{L}$, and suggest no intrinsic dephasing mechanism.}\label{fig1}
\end{figure*}
where the electric field weighted surface-to-volume ratio, $tS^{\mathrm{E}}_{\mathrm{eff}}$/$V^{\mathrm{E}}_{\mathrm{eff}}$, is the surface dielectric participation ratio, $p_{\mathrm{diel}}$$^{8}$.  Of the cavities measured (see Table I), the rectangular TE101 resonance has the most dielectric sensitivity, $p_{\mathrm{diel}}$=2$\times10^{-6}$ for an estimated  surface layer with relative permitivity $\epsilon_{\mathrm{r}}$=10, and thickness $t$=1~nm. While the cylindrical TE011 mode has nominally no electric energy stored at its surfaces, we estimate from finite element simulations that a shape perturbation to the cavity introduces a $p_{\mathrm{diel}}$=4$\times$$10^{-10}$ to the mode. Given the surface dielectric loss properties (t$\sim$1~nm, $\tan\delta$$\leq$10$^{-3}$) inferred from planar resonator measurements\cite{Jonas}, we expect Al 3D cavities could reach $Q$$\approx$$10^{8}$-$10^{12}$.
 
A superconducting resonator can also be damped by a finite surface resistance, $R_{\mathrm{s}}$, perhaps from a finite quasiparticle density. Then, the resonator's quality factor will be given by\cite{Pozar}
\begin{equation}
 Q_{\mathrm{int,H}} = \frac{\omega \mu \lambda}{R_{\mathrm{s}}} \frac{\int_{V}|H|^{2}dV}{\int_{S}|H|^{2}dA\times \lambda} = \frac{Q_{\mathrm{s}} V^{\mathrm{H}}_{\mathrm{eff}}}{\lambda S^{\mathrm{H}}_{\mathrm{eff}}} = \frac{Q_{\mathrm{s}}}{\alpha}~,
\end{equation}
where the magnetic field weighted surface-to-volume ratio, $\lambda S^{\mathrm{H}}_{\mathrm{eff}}$/$V^{\mathrm{H}}_{\mathrm{eff}}$, is the participation ratio, $\alpha$, of the conductor's surface impedance. We recognize $\omega \mu \lambda$ as the resonator's surface reactance, $X_{\mathrm{s}}$. The ratio of $X_{\mathrm{s}}$ to $R_{\mathrm{s}}$ is the resonator's ``surface Q'' - $Q_{\mathrm{s}}$ as reviewed in ref.~[9]. The relationship between $Q$, $Q_{\mathrm{s}}$, and $\alpha$ indicates that the magnetic participation ratio is equivalent to the kinetic inductance fraction. Typical values for $\alpha$ are between $10^{-5}$-$10^{-6}$ for the cavities studied here, many orders of magnitude lower than the (0.005$\sim$0.74) values reported for planar resonators\cite{Day,Leduc}. Therefore aluminum cavity resonators in the quantum regime with at least $Q$$\approx$10$^{9}$, or lifetimes greater than 0.1~s, should be feasible given the same surface resistance values as deposited Al.

\begin{table}[tp]
  \begin{center}
  \caption{Representative Results For Aluminum Cavities - (C) Cylindrical, (R) Rectangular, (e) Acid Treated for 4~Hrs, (*) Overcoupled with $Q_{c}$=3.7M}
    \begin{tabular}{l l l l c c c c} \\
    \hline\hline
   Cavity & Mode & Material & Freq &  $Q_{tot}$ & $Q_{int}$ & $\tau_{int}$  \\
    & & & (GHz) &  ($10^{6}$) &  ($10^{6}$) & (ms)\\
    \hline
    (R1) & TE101 & 6061 &9.513& 2.6 & 5.1 & 0.08	\\ \hline
    (R2)& TE101  &6061(e) & 9.450 & 1.5 & 3.2 & 0.05 \\
     \hline
    (R3) & TE101  &4N &9.464& 4.3 & 5.6 & 0.09	\\
    & & 4N(e)&9.455 & 30& 42 & 0.7 \\
     \hline
    (R4)& TE101  & 5N5 &9.478& 4.2 & 4.8 & 0.08 \\
    \hline		
    (R5)& TE101  & 5N5(e) &9.481 & 40 & 43 & 0.7 \\
    & repeat & 5N5(e) &9.481 & 61 & 69 & 1.2 \\
    \hline
    (C1)&  TE111  & 5N5 &7.690 & 2.4 & 2.4 & 0.06 \\
     & & 5N5(e) &7.700 & 31 & 31 & 0.6 \\
     &   TM111  & 5N5 &11.448 & 1.0 & 1.0 & 0.01 \\
    &  & 5N5(e) &11.448 & 14 & 14 & 0.2 \\
    & TE011 & 5N5 &11.416 & 56 & 150 & 2 \\
     & & 5N5(e)&11.417  & 280 & 609 & 8 \\
     &(*) & 5N5(e)&11.419  & 3.3 & 32 & 0.4 \\
    \hline
    (C2)&  TE011  & 5N5& 11.450 & 15 & 15 & 0.2 \\
        &    & 5N5(e)& 11.440  & 340 & 740 & 10.4 \\
        & repeat   & 5N5(e)& 11.442 & 300 & 520 & 7.2 \\
     \hline		
    \end{tabular}
  \end{center}
\end{table}

Of particular interest to this study is the cylindrical cavity's TE011 mode\cite{Turneaure}, which has nominally zero surface dielectric participation and no currents flowing across the corners of the device. The latter feature allows the cylinder to be sealed at its corners without the dissipation potentially induced at a mechanical joint.  The TE011 mode is explicitly degenerate with the cylindrical TM111 mode. We lift this degeneracy by 30~MHz with a small ring shaped perturbation at the corners of the cavity. Coupling to the TE011 mode is established by loop coupler exciting a sub-cutoff, evanescent mode of a small circular hole (3.5~mm dia) in the top of the cylinder (Fig.1-a). Because our operating frequency is well below the cutoff frequency of this traveling wave, signals are exponentially attenuated in the distance between the coupler and cavity. The evanescent wave approach allows us to consistently reach coupling quality factors in excess of at least a billion, an extent that may prove a challenge with planar-only techniques\cite{Noroozian}. The rectangular waveguide experiments rely on an exposed coaxial center pin coupled through a small hole to the cavity mode's E-field\cite{Paik}, dual to the loop-coupling techniques of the cylinder.

Each cavity is measured in a Cryoperm magnetic shield within a cryogen-free dilution refrigerator. The shunt resonator technique is adapted for our setup (Fig.1-a), where a three port SMA-Tee connector is used to introduce the impedance of the cavity under study to our signal path. In this configuration, both coupling and internal quality factors can be obtained without ambiguity of calibration\cite{Megrant,Geerlings,Khalil}. However, we find that the coupler may add measurable loss to the cavity mode itself in the case of very overcoupled measurements (see asterisked entry in Table I).  We design experiments to be nearly critically coupled for maximum signal to noise at low powers. Microwave signals pass through 20~dB and 30~dB attenuators on the refrigerator's 4~K and 20~mK stages respectively, before entering one port of the SMA-Tee. The second port of the SMA-Tee is connected to two Pamtech isolators at 20~mK and a cryogenic HEMT amplifier at 4~K, which is followed by room temperature amplification and demodulation. The third port of the SMA-Tee is terminated by our coupler and cavity. We analyze the frequency and time domain responses of this circuit at different temperatures and drive powers.

The cavities in our study (Table I) are machined from bulk aluminum with purity ranging from alloy 6061-T6 (95$\%$) to 5N5 (99.9995$\%$) pure. Surfaces are treated in a bath of commercially available phosphoric-nitric acid mix, Transene's Aluminum Etchant Type A at 50~$^{\circ}$C for 4~h, removing 100~$\mu$m of material. The acid bath is refreshed at the 2-h point to avoid saturation. Following acid treatment, the cavities are rinsed with high pressure DI water for 1~min, rinsed with methanol, and blown dry with nitrogen gas. At this point, the purest aluminum samples are high luster with cm-sized grain boundaries. The cavities are then assembled with an indium gasket and stored in room air. No degradation in lifetime has been observed for cavities which remain in such a state for up to six months. Consistent with reports on niobium resonators\cite{Turneaure,Padamsee}, removing 100~$\mu$m of material produces the longest lived resonators in pure, bulk aluminum. Etching as much as 220~$\mu$m shows no signs of further improvement. This surface treatment is not observed to enhance cavities in 6061 alloy. 

Representative results from several variations of cavity type and preparation are shown in Table I. As expected, the TE011 resonance is the longest lived with an intrinsic lifetime of 10.4~ms in one cavity; another nominally identical TE011 resonator reached 8~ms. The longest lived rectangular cavity in this study is 1.2~ms. Negligible variation in cavity properties are observed over long time scales (48~Hrs). Further, the lifetimes extracted for these cavities are observed to be independent of whether phase-sensitive heterodyne measurements (Fig.1-b) or phase-insensitive power-decay techniques (Fig.1-c) are used, which suggests $T^{*}_{2}$$\approx$2$T_{1}$ for these devices.

Surface dielectric loss seems to be excluded as the limitation of our 3D cavity modes from the following observations. First, as described earlier, we would expect to observe significantly higher qualty factors for the same dielectric thickness and loss properties inferred from planar Al resonators. Second, in the cylindrical geometry we measure three modes TE111 ($p_{\mathrm{diel}}$=4$\times$$10^{-7}$), TM111 ($p_{\mathrm{diel}}$=5$\times$$10^{-7}$), and TE011 ($p_{\mathrm{diel}}$=4$\times$$10^{-10}$) with widely varying sensitivity to dielectric loss. Although the quality factors vary, the observed values are inconsistent with any physical value of a surface dielectric loss tangent.

Finally, the lack of power dependence in the quality factors (Fig.1-b) provides further evidence for the irrelevance of surface dielectric loss. Remarkably, we observe no change in both the resonance frequency ($\delta f$$\leq$1~Hz) or the spectral width (FWHM=18$\pm$1~Hz) when increasing the power over ten orders of magnitude from the single photon level, $\bar{n}$=P$_{\mathrm{in}}$$Q$/$\hbar$$\omega^{2}$$\approx1$ (P$_{\mathrm{in}}$=-175~dBm) to (P$_{\mathrm{in}}$=-77~dBm). This is in contrast to planar resonators, where dielectric two-level systems\cite{Martinis} (TLS's) lead to power dependent frequencies and lifetimes.  
\begin{figure}[tp]
\centering
\includegraphics[trim=0cm 1cm 0cm 0cm, clip=false, totalheight=0.4	\textheight]{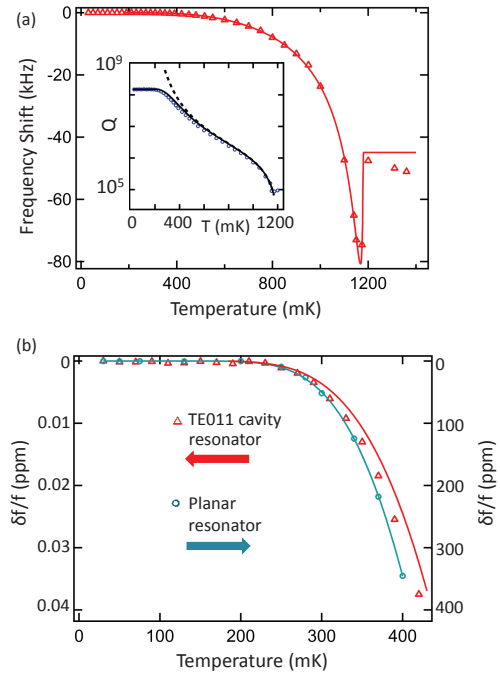}
\caption{(a): The temperature dependence of an Al cavity's quality factor (inset) and frequency (main figure) fit to BCS theory with T$_{\mathrm{c}}$=1.18~K and penetration depth, $\lambda_{0}$=65~nm. (b): Two aluminum resonators (planar and cavity-based) have nearly identical temperature responses, but the cavity resonator's quasiparticle insensitivity is distinguished by its four orders of magnitude smaller frequency shift for equal bath temperatures.}\label{fig2}
\end{figure}
In order to assess the effects of the superconductor's surface impedance on the resonator's performance, we investigate the temperature dependence of the resonator's properties (Fig-2.a). The inverse quality factor (1/Q) and relative frequency shift ($\delta f$/$f$) of the cavity are proportional to the real and imaginary components of its surface impedance by a single proportionality constant, $\alpha$/$\lambda_{0}$, which is a combination of a geometry-dependent factor ($\alpha$) and a materials-dependent factor ($\lambda_{0}$),
\begin{equation}
\frac{1}{Q}+2j\frac{\delta f}{f} = \frac{\alpha}{\omega\mu\lambda_{0}} (R_{s}+j\delta X_{s}).
\end{equation}
As seen in Fig.2-a, the data agree well with the predictions of a numerical integration of Mattis and Bardeen's formulas for AC conductivity of a BCS superconductor\cite{Mattis} with a finite residual quality factor. The sharp drop in frequency (corresponding to a peak in surface reactance) near T$_{\mathrm{c}}$=1.18~K is indicative of pair-breaking due to photon energy equal to the superconducting energy gap. The location of this drop, as well as the plateau in frequency above T$_{\mathrm{c}}$ allows a precise extraction of our aluminum sample's penetration depth, 65$\pm$2~nm. With etching, we find a decrease in penetration depth to 52$\pm$2~nm, indicating a trend toward a cleaner surface and a higher superfluid density. We believe these effects are related to the increased cavity lifetimes following surface treatment. 

The extracted value of alpha ($\alpha$=6$\times$10$^{-6}$) can be compared to the value extracted through the same approach for a planar, quasi-lumped element aluminum resonator on sapphire.  The temperature dependence of $\delta f$/$f$ for the two devices is strikingly similar (Fig. 2-b). However, the size of the shift $\delta f$/$f$ and the conductor participation ratio for the three dimensional resonator is four orders of magnitude smaller, illustrating its dramatically reduced sensitivity to conductor properties.

The origin of the saturation in the quality factors observed at low temperatures (see inset in Fig.2-a) remains an outstanding issue. Similar behavior has been observed in superconducting planar resonators and qubits and could be attributed\cite{Paik,Gianluigi} to a fraction ($x_{qp}$$\sim$10$^{-5}$-10$^{-7}$) of broken Cooper pairs. A similar explanation for the 3D devices presented here however, would require a surpisingly larger quasiparticle density, $x_{qp}$$\sim$10$^{-4}$. Other mechanisms limiting quality factors such as stray magnetic fields, parasitic coupling to the environment, and dissipation associated with current at the seams of the cavity require further investigation.

In conclusion, we have measured several aluminum 3D cavity resonators in the quantum regime. By virtue of their geometry, the surface properties of the materials have a significantly smaller impact on their quality factors, compared to the conventional planar structures. Indeed we see no evidence of the loss due to surface dielectric, and good ageement with the predictions of the Mattis-Bardeen theory for the surface impedance of Al over a wide temperature range. The small participation ratio of the surfaces allowed us to reach quality factors $Q$$\sim$10$^{9}$ and single-photon storage times in excess of 0.01~seconds, and further improvements should be possible\cite{Haroche,Brattke}. The increased lifetimes demonstrated here already make these cavities a valuable resource for quantum information processing with superconducting circuits.

We thank Jean-Michel Raimond, Jonas Zmuidzinas, David Schuster, and Lance Cooley for useful conversations, as well as Dave Johnson for his assistance in prototyping. Facilities use was supported by YINQE and NSF MRSEC DMR 1119826. This research is based upon work supported in part by the Office of the Director of National Intelligence (ODNI), Intelligence Advanced Research Projects Activity (IARPA), via the Army Research Office W911NF-09-1-0369. The views and conclusions contained herein are those of the authors and should not be interpreted as necessarily representing the official policies or endorsements, either expressed or implied, of ODNI, IARPA, or the U.S. Government. The U.S. Government is authorized to reproduce and distribute reprints for Governmental purposes notwithstanding any copyright annotation thereon.

%


\bibliographystyle{apsrev}

\begin{thebibliography}{20}
\expandafter\ifx\csname natexlab\endcsname\relax\def\natexlab#1{#1}\fi
\expandafter\ifx\csname bibnamefont\endcsname\relax
  \def\bibnamefont#1{#1}\fi
\expandafter\ifx\csname bibfnamefont\endcsname\relax
  \def\bibfnamefont#1{#1}\fi
\expandafter\ifx\csname citenamefont\endcsname\relax
  \def\citenamefont#1{#1}\fi
\expandafter\ifx\csname url\endcsname\relax
  \def\url#1{\texttt{#1}}\fi
\expandafter\ifx\csname urlprefix\endcsname\relax\def\urlprefix{URL }\fi
\providecommand{\bibinfo}[2]{#2}
\providecommand{\eprint}[2][]{\url{#2}}

\bibitem[{\citenamefont{Houck et~al.}(2008)\citenamefont{Houck, Schreier,
  Johnson, Chow, Koch, Gambetta, Schuster, Frunzio, Devoret, and
  Girvin}}]{Houck}
\bibinfo{author}{\bibfnamefont{A.~A.} \bibnamefont{Houck}},
  \bibinfo{author}{\bibfnamefont{J.~A.} \bibnamefont{Schreier}},
  \bibinfo{author}{\bibfnamefont{B.~R.} \bibnamefont{Johnson}},
  \bibinfo{author}{\bibfnamefont{J.~M.} \bibnamefont{Chow}},
  \bibinfo{author}{\bibfnamefont{J.}~\bibnamefont{Koch}},
  \bibinfo{author}{\bibfnamefont{J.~M.} \bibnamefont{Gambetta}},
  \bibinfo{author}{\bibfnamefont{D.~I.} \bibnamefont{Schuster}},
  \bibinfo{author}{\bibfnamefont{L.}~\bibnamefont{Frunzio}},
  \bibinfo{author}{\bibfnamefont{M.~H.} \bibnamefont{Devoret}},
  \bibinfo{author}{\bibfnamefont{S.~M.}  \bibnamefont{Girvin}}, 
    \bibnamefont{and}
    \bibinfo{author}{\bibnamefont{R.~J.} \bibnamefont{Schoelkopf}},
  \bibinfo{journal}{Physical Review Letters}
  \textbf{\bibinfo{volume}{101}}, \bibinfo{pages}{80502}
  (\bibinfo{year}{2008}).

\bibitem[{\citenamefont{Johnson et~al.}(2010)\citenamefont{Johnson, Reed,
  Houck, Schuster, Bishop, Ginossar, Gambetta, DiCarlo, Frunzio, Girvin
  et~al.}}]{Johnson}
\bibinfo{author}{\bibfnamefont{B.~R.} \bibnamefont{Johnson}},
  \bibinfo{author}{\bibfnamefont{M.~D.} \bibnamefont{Reed}},
  \bibinfo{author}{\bibfnamefont{A.~A.} \bibnamefont{Houck}},
  \bibinfo{author}{\bibfnamefont{D.~I.} \bibnamefont{Schuster}},
  \bibinfo{author}{\bibfnamefont{L.~S.}~\bibnamefont{Bishop}},
  \bibinfo{author}{\bibfnamefont{E.} \bibnamefont{Ginossar}},
  \bibinfo{author}{\bibfnamefont{J.~M.} \bibnamefont{Gambetta}},
  \bibinfo{author}{\bibfnamefont{L.} \bibnamefont{DiCarlo}},
  \bibinfo{author}{\bibfnamefont{L.} \bibnamefont{Frunzio}},
  \bibinfo{author}{\bibfnamefont{S.~M.} \bibnamefont{Girvin}},
  \bibnamefont{and} \bibinfo{author}{\bibnamefont{R.~J.} \bibnamefont{Schoelkopf}},
  \bibinfo{journal}{Nature Physics}
  \textbf{\bibinfo{volume}{6}}, \bibinfo{pages}{663} (\bibinfo{year}{2010}).

\bibitem[{\citenamefont{Hatridge et~al.}(2013)\citenamefont{Hatridge, Shankar,
  Mirrahimi, Schackert, Geerlings, Brecht, Sliwa, Abdo, Frunzio, Girvin
  et~al.}}]{Hatridge}
\bibinfo{author}{\bibfnamefont{M.}~\bibnamefont{Hatridge}},
  \bibinfo{author}{\bibfnamefont{S.}~\bibnamefont{Shankar}},
  \bibinfo{author}{\bibfnamefont{M.}~\bibnamefont{Mirrahimi}},
  \bibinfo{author}{\bibfnamefont{F.}~\bibnamefont{Schackert}},
  \bibinfo{author}{\bibfnamefont{K.}~\bibnamefont{Geerlings}},
  \bibinfo{author}{\bibfnamefont{T.}~\bibnamefont{Brecht}},
  \bibinfo{author}{\bibfnamefont{K.~M.} \bibnamefont{Sliwa}},
  \bibinfo{author}{\bibfnamefont{B.}~\bibnamefont{Abdo}},
  \bibinfo{author}{\bibfnamefont{L.}~\bibnamefont{Frunzio}},
  \bibinfo{author}{\bibfnamefont{S.~M.} \bibnamefont{Girvin}},
  \bibinfo{author}{\bibfnamefont{R.~J.} \bibnamefont{Schoelkopf}},
  \bibnamefont{and} \bibinfo{author}{\bibnamefont{M.~H.} \bibnamefont{Devoret}},
  \bibinfo{journal}{Science}
  \textbf{\bibinfo{volume}{339}}, \bibinfo{pages}{178} (\bibinfo{year}{2013}).

\bibitem[{\citenamefont{Mariantoni et~al.}(2011)\citenamefont{Mariantoni, Wang,
  Yamamoto, Neeley, Bialczak, Chen, Lenander, Lucero, O'Connell, Sank, et~al.}}]{Mariantoni}
\bibinfo{author}{\bibfnamefont{M.}~\bibnamefont{Mariantoni}},
  \bibinfo{author}{\bibfnamefont{H.}~\bibnamefont{Wang}},
  \bibinfo{author}{\bibfnamefont{T.}~\bibnamefont{Yamamoto}},
  \bibinfo{author}{\bibfnamefont{M.}~\bibnamefont{Neeley}},
  \bibinfo{author}{\bibfnamefont{R.~C.} \bibnamefont{Bialczak}},
  \bibinfo{author}{\bibfnamefont{Y.}~\bibnamefont{Chen}},
  \bibinfo{author}{\bibfnamefont{M.}~\bibnamefont{Lenander}},
  \bibinfo{author}{\bibfnamefont{E.}~\bibnamefont{Lucero}},
  \bibinfo{author}{\bibfnamefont{A.}~\bibnamefont{O'Connell}},
  \bibinfo{author}{\bibfnamefont{D.}~\bibnamefont{Sank}}  {\bibfnamefont{et~al.}},
  \bibinfo{journal}{Science} \textbf{\bibinfo{volume}{334}},
  \bibinfo{pages}{61} (\bibinfo{year}{2011}).

\bibitem[{\citenamefont{Megrant et~al.}(2012)\citenamefont{Megrant, Neill,
  Barends, Chiaro, Chen, Feigl, Kelly, Lucero, Mariantoni, O'Malley
  et~al.}}]{Megrant}
\bibinfo{author}{\bibfnamefont{A.}~\bibnamefont{Megrant}},
  \bibinfo{author}{\bibfnamefont{C.}~\bibnamefont{Neill}},
  \bibinfo{author}{\bibfnamefont{R.}~\bibnamefont{Barends}},
  \bibinfo{author}{\bibfnamefont{B.}~\bibnamefont{Chiaro}},
  \bibinfo{author}{\bibfnamefont{Y.}~\bibnamefont{Chen}},
  \bibinfo{author}{\bibfnamefont{L.}~\bibnamefont{Feigl}},
  \bibinfo{author}{\bibfnamefont{J.}~\bibnamefont{Kelly}},
  \bibinfo{author}{\bibfnamefont{E.}~\bibnamefont{Lucero}},
  \bibinfo{author}{\bibfnamefont{M.}~\bibnamefont{Mariantoni}},
  \bibinfo{author}{\bibfnamefont{P.~J.~J.} \bibnamefont{O'Malley}} {\bibfnamefont{et~al.}},
  \bibnamefont{and} \bibinfo{author}{\bibfnamefont{A.~N.}~\bibnamefont{Cleland}},
  \bibinfo{journal}{Applied Physics Letters}
  \textbf{\bibinfo{volume}{100}}, \bibinfo{pages}{113510}
  (\bibinfo{year}{2012}).

\bibitem[{\citenamefont{Barends et~al.}(2010)\citenamefont{Barends,
  Vercruyssen, Endo, de~Visser, Zijlstra, Klapwijk, Diener, Yates, Baselmans}}]{Barends}
\bibinfo{author}{\bibfnamefont{R.}~\bibnamefont{Barends}},
  \bibinfo{author}{\bibfnamefont{N.}~\bibnamefont{Vercruyssen}},
  \bibinfo{author}{\bibfnamefont{A.}~\bibnamefont{Endo}}, 
  \bibinfo{author}{\bibfnamefont{P.~J.} \bibnamefont{de~Visser}},
  \bibinfo{author}{\bibfnamefont{T.}~\bibnamefont{Zijlstra}},
  \bibinfo{author}{\bibfnamefont{T.~M.}~\bibnamefont{Klapwijk}}, 
  \bibinfo{author}{\bibfnamefont{P.}~\bibnamefont{Diener}}, 
  \bibinfo{author}{\bibfnamefont{S.~J.~C.} \bibnamefont{Yates}},
  \bibnamefont{and} \bibinfo{author}{\bibfnamefont{J.~J.~A.} \bibnamefont{Baselmans}},
  \bibinfo{journal}{Applied Physics Letters} \textbf{\bibinfo{volume}{97}},
  \bibinfo{pages}{023508} (\bibinfo{year}{2010}).

\bibitem[{\citenamefont{Geerlings et~al.}(2012)\citenamefont{Geerlings,
  Shankar, Edwards, Frunzio, Schoelkopf, and Devoret}}]{Geerlings}
\bibinfo{author}{\bibfnamefont{K.}~\bibnamefont{Geerlings}},
  \bibinfo{author}{\bibfnamefont{S.}~\bibnamefont{Shankar}},
  \bibinfo{author}{\bibfnamefont{E.}~\bibnamefont{Edwards}},
  \bibinfo{author}{\bibfnamefont{L.}~\bibnamefont{Frunzio}},
  \bibinfo{author}{\bibfnamefont{R.~J.} \bibnamefont{Schoelkopf}},
  \bibnamefont{and} \bibinfo{author}{\bibfnamefont{M.~H.}
  \bibnamefont{Devoret}}, \bibinfo{journal}{Applied Physics Letters}
  \textbf{\bibinfo{volume}{100}}, \bibinfo{pages}{192601}
  (\bibinfo{year}{2012}).

\bibitem[{\citenamefont{Gao et~al.}(2008)\citenamefont{Gao, Daal, Vayonakis,
  Kumar, Zmuidzinas, Sadoulet, Mazin, Day, and Leduc}}]{Gao}
\bibinfo{author}{\bibfnamefont{J.}~\bibnamefont{Gao}},
  \bibinfo{author}{\bibfnamefont{M.}~\bibnamefont{Daal}},
  \bibinfo{author}{\bibfnamefont{A.}~\bibnamefont{Vayonakis}},
  \bibinfo{author}{\bibfnamefont{S.}~\bibnamefont{Kumar}},
  \bibinfo{author}{\bibfnamefont{J.}~\bibnamefont{Zmuidzinas}},
  \bibinfo{author}{\bibfnamefont{B.}~\bibnamefont{Sadoulet}},
  \bibinfo{author}{\bibfnamefont{B.~A.} \bibnamefont{Mazin}},
  \bibinfo{author}{\bibfnamefont{P.~K.} \bibnamefont{Day}}, 
  \bibnamefont{and} \bibinfo{author}{\bibfnamefont{H.~G.} \bibnamefont{Leduc}},
  \bibinfo{journal}{Applied Physics Letters} \textbf{\bibinfo{volume}{92}},
  \bibinfo{pages}{152505} (\bibinfo{year}{2008}).

\bibitem[{\citenamefont{Zmuidzinas}(2012)}]{Jonas}
\bibinfo{author}{\bibfnamefont{J.}~\bibnamefont{Zmuidzinas}},
  \bibinfo{journal}{Annual Review of Condensed Matter Physics}
  \textbf{\bibinfo{volume}{3}}, \bibinfo{pages}{169} (\bibinfo{year}{2012}).

\bibitem[{\citenamefont{Turneaure}(1968)}]{Turneaure}
\bibinfo{author}{\bibfnamefont{J.~P.} \bibnamefont{Turneaure}},
\bibnamefont{and} \bibinfo{author}{\bibfnamefont{I.} \bibnamefont{Weissman}},
  \bibinfo{journal}{Journal of Applied Physics} \textbf{\bibinfo{volume}{39}},
  \bibinfo{pages}{4417} (\bibinfo{year}{1968}).

\bibitem[{\citenamefont{Padamsee}(2001)}]{Padamsee}
\bibinfo{author}{\bibfnamefont{H.}~\bibnamefont{Padamsee}},
  \bibinfo{journal}{Superconductor Science and Technology}
  \textbf{\bibinfo{volume}{14}}, \bibinfo{pages}{R28} (\bibinfo{year}{2001}).

\bibitem[{\citenamefont{Kuhr et~al.}(2007)\citenamefont{Kuhr, Gleyzes, Guerlin,
  Bernu, Hoff, Del{\'e}glise, Osnaghi, Brune, Raimond, Haroche
  et~al.}}]{Haroche}
\bibinfo{author}{\bibfnamefont{S.}~\bibnamefont{Kuhr}},
  \bibinfo{author}{\bibfnamefont{S.}~\bibnamefont{Gleyzes}},
  \bibinfo{author}{\bibfnamefont{C.}~\bibnamefont{Guerlin}},
  \bibinfo{author}{\bibfnamefont{J.}~\bibnamefont{Bernu}},
  \bibinfo{author}{\bibfnamefont{U.~B.} \bibnamefont{Hoff}},
  \bibinfo{author}{\bibfnamefont{S.}~\bibnamefont{Del{\'e}glise}},
  \bibinfo{author}{\bibfnamefont{S.}~\bibnamefont{Osnaghi}},
  \bibinfo{author}{\bibfnamefont{M.}~\bibnamefont{Brune}},
  \bibinfo{author}{\bibfnamefont{J.~M.} \bibnamefont{Raimond}},
  \bibinfo{author}{\bibfnamefont{S.}~\bibnamefont{Haroche}}  {\bibfnamefont{et~al.}},
  \bibnamefont{et~al.}, \bibinfo{journal}{Applied Physics Letters}
  \textbf{\bibinfo{volume}{90}}, \bibinfo{pages}{164101}
  (\bibinfo{year}{2007}).

\bibitem[{\citenamefont{Brattke et~al.}(2001)\citenamefont{Brattke, Varcoe, and
  Walther}}]{Brattke}
\bibinfo{author}{\bibfnamefont{S.}~\bibnamefont{Brattke}},
  \bibinfo{author}{\bibfnamefont{B.}~\bibnamefont{Varcoe}}, \bibnamefont{and}
  \bibinfo{author}{\bibfnamefont{H.}~\bibnamefont{Walther}},
  \bibinfo{journal}{Physical Review Letters} \textbf{\bibinfo{volume}{86}},
  \bibinfo{pages}{3534} (\bibinfo{year}{2001}).

\bibitem[{\citenamefont{Paik et~al.}(2011)\citenamefont{Paik, Schuster, Bishop,
  Kirchmair, Catelani, Sears, Johnson, Reagor, Frunzio, Glazman, et~al.}}]{Paik}
\bibinfo{author}{\bibfnamefont{H.}~\bibnamefont{Paik}},
  \bibinfo{author}{\bibfnamefont{D.}~\bibnamefont{Schuster}},
  \bibinfo{author}{\bibfnamefont{L.}~\bibnamefont{Bishop}},
  \bibinfo{author}{\bibfnamefont{G.}~\bibnamefont{Kirchmair}},
  \bibinfo{author}{\bibfnamefont{G.}~\bibnamefont{Catelani}},
  \bibinfo{author}{\bibfnamefont{A.}~\bibnamefont{Sears}},
  \bibinfo{author}{\bibfnamefont{B.}~\bibnamefont{Johnson}},
  \bibinfo{author}{\bibfnamefont{M.}~\bibnamefont{Reagor}},
  \bibinfo{author}{\bibfnamefont{L.}~\bibnamefont{Frunzio}},
  \bibinfo{author}{\bibfnamefont{L.}~\bibnamefont{Glazman}} {\bibfnamefont{et~al.}},
  \bibinfo{journal}{Physical Review Letters} \textbf{\bibinfo{volume}{107}},
  \bibinfo{pages}{240501} (\bibinfo{year}{2011}).

\bibitem[{\citenamefont{Pozar}(2005)}]{Pozar}
\bibinfo{author}{\bibfnamefont{D.}~\bibnamefont{Pozar}},
  \emph{\bibinfo{title}{{Microwave Engineering}}}, \bibinfo{number}{3$^{rd}$ Edition},
  \bibinfo{publisher}{John Wiley $\&$ Sons, Inc.}, (\bibinfo{year}{2005}).

\bibitem[{\citenamefont{Day et~al.}(2003)\citenamefont{Day, Leduc, Mazin,
  Vayonakis, and Zmuidzinas}}]{Day}
\bibinfo{author}{\bibfnamefont{P.~K.} \bibnamefont{Day}},
  \bibinfo{author}{\bibfnamefont{H.~G.} \bibnamefont{Leduc}},
  \bibinfo{author}{\bibfnamefont{B.~A.} \bibnamefont{Mazin}},
  \bibinfo{author}{\bibfnamefont{A.}~\bibnamefont{Vayonakis}},
  \bibnamefont{and} \bibinfo{author}{\bibfnamefont{J.}~\bibnamefont{Zmuidzinas}},
  \bibinfo{journal}{Nature} \textbf{\bibinfo{volume}{425}},
  \bibinfo{pages}{817} (\bibinfo{year}{2003}).

\bibitem[{\citenamefont{Leduc et~al.}(2010)\citenamefont{Leduc, Bumble, Day,
  Eom, Gao, Golwala, Mazin, McHugh, Merrill, Moore et~al.}}]{Leduc}
\bibinfo{author}{\bibfnamefont{H.~G.} \bibnamefont{Leduc}},
  \bibinfo{author}{\bibfnamefont{B.}~\bibnamefont{Bumble}},
  \bibinfo{author}{\bibfnamefont{P.~K.} \bibnamefont{Day}},
  \bibinfo{author}{\bibfnamefont{B.~H.} \bibnamefont{Eom}},
  \bibinfo{author}{\bibfnamefont{J.}~\bibnamefont{Gao}},
  \bibinfo{author}{\bibfnamefont{S.}~\bibnamefont{Golwala}},
  \bibinfo{author}{\bibfnamefont{B.~A.} \bibnamefont{Mazin}},
  \bibinfo{author}{\bibfnamefont{S.}~\bibnamefont{McHugh}},
  \bibinfo{author}{\bibfnamefont{A.}~\bibnamefont{Merrill}},
  \bibinfo{author}{\bibfnamefont{D.~C.} \bibnamefont{Moore}} {\bibfnamefont{et~al.}},
  \bibinfo{journal}{Applied Physics Letters}
  \textbf{\bibinfo{volume}{97}}, \bibinfo{pages}{102509}
  (\bibinfo{year}{2010}).

\bibitem[{\citenamefont{Martinis et~al.}(2005)\citenamefont{Martinis, Cooper, McDermott,
  Steffen, Ansman, Osborn, Cicak, Oh, Pappas, Yu}}]{Martinis}
\bibinfo{author}{\bibfnamefont{John.~M.} \bibnamefont{Martinis}},
  \bibinfo{author}{\bibfnamefont{K.~B.}~\bibnamefont{Cooper}},
  \bibinfo{author}{\bibfnamefont{R.} \bibnamefont{McDermott}},
  \bibinfo{author}{\bibfnamefont{M.} \bibnamefont{Steffan}},
  \bibinfo{author}{\bibfnamefont{M.}~\bibnamefont{Ansmann}},
  \bibinfo{author}{\bibfnamefont{K.}~\bibnamefont{Osborn}},
  \bibinfo{author}{\bibfnamefont{K.} \bibnamefont{Cicak}},
  \bibinfo{author}{\bibfnamefont{S.}~\bibnamefont{Oh}},
  \bibinfo{author}{\bibfnamefont{D.~P.}~\bibnamefont{Pappas}},
  \bibinfo{author}{\bibfnamefont{R.~W.} \bibnamefont{Simmonds}},
  \bibnamefont{and}  \bibinfo{author}{\bibfnamefont{C.~C.} \bibnamefont{Yu}},
  \bibinfo{journal}{Physical Review Letters}
  \textbf{\bibinfo{volume}{95}}, \bibinfo{pages}{210503}
  (\bibinfo{year}{2005}).

\bibitem[{\citenamefont{Noroozian et~al.}(2012)\citenamefont{Noroozian, Day, Eom,
  Leduc, Yu, Zmuidzinas}}]{Noroozian}
\bibinfo{author}{\bibfnamefont{O.} \bibnamefont{Noroozian}},
  \bibinfo{author}{\bibfnamefont{P.}~\bibnamefont{Day}},
  \bibinfo{author}{\bibfnamefont{B.~H.} \bibnamefont{Eom}},
  \bibinfo{author}{\bibfnamefont{H.} \bibnamefont{Leduc}},
  \bibnamefont{and}  \bibinfo{author}{\bibfnamefont{J.} \bibnamefont{Zmuidzinas}},
  \bibinfo{journal}{IEEE
Trans. Microwave Theory Tech.}
  \textbf{\bibinfo{volume}{60}}, \bibinfo{pages}{1235}
  (\bibinfo{year}{2012}). 

  

\bibitem[{\citenamefont{Khalil et~al.}(2012)\citenamefont{Khalil, Stoutimore,
  Wellstood, and Osborn}}]{Khalil}
\bibinfo{author}{\bibfnamefont{M.~S.} \bibnamefont{Khalil}},
  \bibinfo{author}{\bibfnamefont{M.~J.~A.} \bibnamefont{Stoutimore}},
  \bibinfo{author}{\bibfnamefont{F.~C.} \bibnamefont{Wellstood}},
  \bibnamefont{and} \bibinfo{author}{\bibfnamefont{K.~D.}
  \bibnamefont{Osborn}}, \bibinfo{journal}{Journal of Applied Physics}
  \textbf{\bibinfo{volume}{111}}, \bibinfo{pages}{054510}
  (\bibinfo{year}{2012}).
  
\bibitem[{\citenamefont{Mattis and Bardeen}(1958)\citenamefont{Mattis and Bardeen}}]{Mattis}
\bibinfo{author}{\bibfnamefont{D.~C.} \bibnamefont{Mattis}},
  \bibnamefont{and} \bibinfo{author}{\bibfnamefont{J.}
  \bibnamefont{Bardeen}}, \bibinfo{journal}{Physical Review}
  \textbf{\bibinfo{volume}{111}}, \bibinfo{pages}{412}
  (\bibinfo{year}{1958}).
  
\bibitem[{\citenamefont{Catelani et~al.}(2012)\citenamefont{Catelani, Schoelkopf, Devoret, and Glazman}}]{Gianluigi}
\bibinfo{author}{\bibfnamefont{G.} \bibnamefont{Catelani}},
\bibinfo{author}{\bibfnamefont{R.~J.} \bibnamefont{Schoelkopf}},
\bibinfo{author}{\bibfnamefont{M.~H.} \bibnamefont{Devoret}},
  \bibnamefont{and} \bibinfo{author}{\bibfnamefont{L.~I.}
  \bibnamefont{Glazman}}, \bibinfo{journal}{Physical Review B}
  \textbf{\bibinfo{volume}{84}}, \bibinfo{pages}{064517}
  (\bibinfo{year}{2011}).
  
\end{thebibliography}

\end{document}